\documentclass{revtex4}
\usepackage{amsthm,amssymb,amsmath}				
\usepackage{graphicx,comment}
\usepackage{float}   
 \usepackage{xcolor}

\begin{document}

\title{Dynamics of the Modified Emden and pseudo-Modified Emden Equations: position-dependent mass, invariance and exact solvability}
\author{Omar Mustafa}
\email{omar.mustafa@emu.edu.tr}
\affiliation{Department of Physics, Eastern Mediterranean University, G. Magusa, north
Cyprus, Mersin 10 - Turkey,\\
Tel.: +90 392 6301378; fax: +90 3692 365 1604.}

\begin{abstract}
\textbf{Abstract:}\ We consider the modified Emden equation (MEE) and introduce its most general solution, using the most general solution for the simple harmonic oscillator's linear dynamical equation (i.e., the initial conditions shall be identified by the PDM-MEE problem at hand) . We use a general nonlocal point transformation and show that modified Emden dynamical equation is transformed to describe position-dependent mass (PDM) classical particles. Two PDM-MEE-type classical particles are used as illustrative examples, and their exact solutions are reported. Under specific parametric considerations, the phase-space trajectories are reported for the MEE-type and for PDM-MEE-type classical particles.


\end{abstract}

\maketitle

\section{Introduction}

A classical particle, with mass $m_{\circ }$ ($m_{\circ }=1$ throughout unless otherwise mentioned), moving under the influence of  a conservative quartic anharmonic potential $V\left( u\right) = \frac{1}{2}%
\omega ^{2}u^{2}+\frac{1}{4}\beta ^{2}u^{4}$ is described by the standard Lagrangian 
\begin{equation}
L\left( u,\dot{u},t\right) =\frac{1}{2}\dot{u}^{2}-\left[ \frac{1}{2}\omega
^{2}u^{2}+\frac{1}{4}\beta ^{2}u^{4}\right] ,  \label{q-Lagrangian1}
\end{equation}%
(i.e., $L=T-V$). Moreover, if this particle is subjected to a non-conservative dissipative Rayleigh force field $\mathcal{R}\left( u,\dot{u%
}\right) =\frac{1}{2}\alpha u\dot{u}^{2}$ , then the corresponding Euler-Lagrange dynamical equation reads%
\begin{equation}
\frac{d}{dt}\left( \frac{\partial L}{\partial \dot{u}}\right) -\frac{%
\partial L}{\partial u}+\frac{\partial \mathcal{R}}{\partial \dot{u}}%
=0\Longleftrightarrow \ddot{u}+\alpha \,u\,\dot{u}+\omega ^{2}\,u+\beta
^{2}u^{3}=0.  \label{q-MEE}
\end{equation}%
This equation is known as the modified Emden equation (MEE), which describes a cubic anharmonic oscillator with a linear forcing term $\omega ^{2}u\left(
t\right) $ and an additional position-dependent nonlinear-damping force $\alpha \,u(t)\,\dot{u}(t)$ (e.g., \cite{Chandrasekar 2006,Chandrasekar 2012} ). Where $\alpha ,\omega >0$ and $\alpha $ is the frictional coefficient, $%
\omega =\sqrt{k/m_{\circ }}$ is the frequency of the standard simple harmonic oscillator, with the force constant $k>0$, and $\beta $ is a coupling constant.  Such a nonlinear differential equation finds its applicability in different fields of study. Amongst are, the equilibrium configuration of spherical gas clouds under mutual attraction of their molecules and admit the thermodynamical laws \cite{Leach 1985}, the spherically symmetric expansion or collapse of relativistically gravitating mass \cite{Mc Vittie 1980}, the one-dimensional analog of the Yang-Mills boson gauge theory \cite{Yang-Mills 1954}, etc. Throughout the current proposal, we shall refer to it as MEE, in short.

The MEE, nevertheless, admits an exact solution \cite{Chandrasekar 2012,Chandrasekar-PRE 2005} in the form of%
\begin{equation}
u\left( t\right) =\frac{C\sin \left( \omega t+\varphi \right) }{1-\eta\, C\cos
\left( \omega t+\varphi \right) }\,;\,\,\,0\leq \text{ }\eta <\frac{1}{C}.
\label{q-MEE-solution}
\end{equation}%
Where $\eta =\alpha /3\,m_{\circ }\,\omega \geq 0$ is  a damping related parameter,  and $0<C<1/\eta$ is the amplitude. Then, the corresponding canonical momentum is simply given by%
\begin{equation}
 P(t)=\frac{\partial L}{\partial \dot{u}}=\dot{u}(t).\label{q-momentum} 
\end{equation}%
When Eq. (\ref{q-MEE-solution}) is substituted in Eq. (\ref{q-MEE}), it implies that $\alpha =3\,\omega\, \eta $, $\beta ^{2}=\omega ^{2}\eta
^{2}\longrightarrow \beta =\omega\, \eta $.  As such, equation (\ref{q-MEE}) can be rewritten as%
\begin{equation}
\ddot{u}\left( t\right) +3\,\omega\, \eta \,u\left( t\right) \,\dot{u}\left(
t\right) +\omega ^{2}u\left( t\right) +\omega ^{2}\eta ^{2}u\left( t\right)
^{3}=0,  \label{q-MEE1}
\end{equation}%
Obviously, the parametric structure of each term identifies the dynamical correlation between the forces involved in the problem at hand. Yet, the solution in Eq. (\ref{q-MEE-solution}) is an immediate consequence of the linearization of Eq. (\ref{q-MEE}) into a linear harmonic oscillator 
\begin{equation}
\ddot{U}+\omega ^{2}U=0\,;\, \, U\left( t\right) = A\sin \left( \omega t+\varphi
\right) ,  \label{HO-solution}
\end{equation}%
using a nonlocal transformation of the form 
\begin{equation}
U\left( t\right) =u\left( t\right) \exp \left( \beta \int u\left( t\right)
dt\right)   \label{PT1}
\end{equation}%
(for more details on this issue the reader may refer to Chandrasekar and coworkers \cite{Chandrasekar 2006,Chandrasekar 2012,Chandrasekar-PRE 2005}). It is clear that, the solution of the typical harmonic oscillator $U\left( t\right) =A\sin \left( \omega
t+\varphi \right) $ belongs to the initial conditions $U\left( 0\right)
=0$ and $\dot{U}\left( 0\right) \neq 0$ at $\varphi=0$ . Moreover, the dynamical equation (\ref{q-MEE1}) is a second-order autonomous differential equation and is known to exhibit certain unusual nonlinear properties (e.g., \cite{Chandrasekar 2006} ). The frequency of oscillation $\omega $, for example, is an amplitude-independent one (i.e., isochronic oscillations similar to the linear harmonic oscillator) documenting in effect that an amplitude-dependent frequency is not always a consequential property of the nonlinear dynamical systems.

On the other hand, it has been asserted that a point mass moving within the curved space transforms into an effective position-dependent mass (PDM) in Euclidean space \cite{Khlevniuk 2018}. Exploring/studying the effect of such coordinate transformation (that manifestly introduces PDM metaphoric concept) on some dynamical systems deserves some attention, and is the main objective of our methodical proposal, therefore.  This PDM concept has, in fact, inspired research activities on PDM (both in classical and quantum mechanical models) ever since the introduction of the prominent Mathews-Lakshmanan oscillator back in 1974 \cite{M-L 1974}. For example, a point particle moving within a specific set of coordinates may very well be transformed into an effectively PDM particle in the new deformed coordinates (c.f., e.g., the sample of references \cite{Khlevniuk 2018,Carinena Ranada Sant 2004,Mustafa 2015,Mustafa 2019,Mustafa Algadhi 2019,Mustafa Habib 2007,Mustafa 2020}). Such research activities/studies have used both non-standard Lagrangians/Hamiltonians (c.f., e.g.,  \cite{M-L 1974,Mustafa 2015,Quesne 2015,Tiwari 2013,Lak-Chand 2013,Pradeep
2009,Chand-Lak 2007,Musielak 2008,Bhuvaneswari 2012,Carinena Ranada 2005,Mustafa Phys.Scr. 2020,Ranada 2016,Carinena Herranz 2017,Bagchi Ghosh
2013,Mustafa 2013} and standard Lagrangians/Hamiltonians. Whilst the intimate long-standing \emph{gain-loss correlation} between the kinetic and
potential energies is ignored in the non-standard Lagrangians/Hamiltonians, it has been kept intact in the standard Lagrangians/Hamiltonians (i.e., $L=T-V$ and $H=T+V$), so that the total energy of the dynamical system is an integral of motion \cite{Carl Bender 2016,Mustafa 2020,Mustafa arXiv}. 

In classical mechanics, a PDM-particle with $M(x)=m_\circ\,G(x)$ (where $G(x)$ is a positive valued dimensionless scalar multiplier) is shown to have a canonical momentum $p(x)=M(x)\,\dot{x}=m_\circ\,G(x)\,\dot{x}$ for a PDM-Lagrangian $\mathcal{L}=M(x)\,\dot{x}^2 /2-V(x)$. In quantum mechanics, nevertheless, the focus was on the prominent von Roos PDM-Hamiltonian \cite{von Roos,Dutra Almeida 2000,dos Santos 2021,Nabulsi1 2020,Nabulsi2 2020,Mustafa Habib1 2006,Bagchi 2005,Mustafa 2011,Cruz 2009,Habib Mustafa 2013,Quesne 2004,Quesne 2019,da Costa1 2020}, where the PDM-operator \cite{Mustafa Algadhi 2019} is shown to be given  (in $\hbar=2m_\circ=1$ units) by%
\begin{equation*}
\hat{p}(x)=-i\left(\partial_{x}-\frac{\partial_{x}G(x)}{4G(x)}\right).
\end{equation*}%
Which, in turn, corresponds to the PDM kinetic energy operator%
\begin{equation*}
\hat{T}(x)=-G(x)^{-1/4}\,\partial_{x}\,G(x)^{-1/2}\,\partial_{x}\,G(x)^{-1/4},
\end{equation*}%
that is known in the literature as Mustafa and Mazharimousavi's ordering \cite{Mustafa Habib 2007,Mustafa 2020}, and belongs the the set of von Roos kinetic energy operators \cite{von Roos}. It would be interesting, therefore, to study the modified Emden equation (\ref{q-MEE1}) for PDM particles, which is in fact the focal point of the current methodical proposal. Throughout, we shall identify a classical-state by $\{x(t),p(t)\}$ \cite{R Shankar} which corresponds to a specific phase-space trajectory. To the best of our knowledge, the current study has never been reported elsewhere.

In this article, we recollect (in section 2) the most general solution for the harmonic oscillator in Eq. (\ref{HO-solution}) and map it report the most general solution for the MEE (\ref{q-MEE}), where the initial conditions are to be identified for the corresponding PDM systems at hand. In the same section, we report some MEE classical-states at some specific parametric structures. In section 3, we introduce the PDM-settings for the MEE Eq. (\ref{q-MEE}) and show that the dynamical equations for PDM-MEE systems are invariant with that of  Eq. (\ref{q-MEE}) and, therefore, are quasi-MEE that inherit the solutions of MEE. Two PDM quasi-MEE classical particles are used as illustrative examples. We give our concluding remarks in section 4.

\section{MEE solution revisited and generalized}

Let us consider a more general nonlocal point transformation%
\begin{equation}
U=u\left( t\right) \exp \left( \int \left[ \alpha g\left( t\right) +\beta
f\left( t\right) \right] dt\right) ;\, \, g\left( t\right) ,f\left( t\right) \in 
\mathbb{R}
,  \label{PT2-gen}
\end{equation}
for the harmonic oscillator of  Eq. (\ref{HO-solution}), where $g\left( t\right) $, and $f\left( t\right) $ are two arbitrary real functions, and $\alpha $ and $%
\beta $ are as defined in  Eq. (\ref{q-MEE}) (for more details on the nonlocal point transformation, the reader may refer to, e.g., \cite{ Muriel 2010,Demitry 2020} and related references cited therein).  This assumption, when substituted in Eq. (\ref{HO-solution}), would yield%
\begin{equation}
\ddot{u}\left( t\right) +\dot{u}\left( t\right) \left[ 2\alpha g\left(
t\right) +2\beta f\left( t\right) \right] +\,u\left( t\right) \,\left[
\alpha \dot{g}\left( t\right) +\beta \dot{f}\left( t\right) \right] +u\left(
t\right) \,\left[ \alpha g\left( t\right) +\beta f\left( t\right) \right]
^{2}+\omega ^{2}u\left( t\right) =0.  \label{MEE3}
\end{equation}%
Comparing this equation with the MEE in Eq. (\ref{q-MEE}), the parametric correlations arise in the process so that%
\begin{equation}
2\alpha g\left( t\right) +2\beta f\left( t\right) =\frac{2}{3}\alpha u\left(
t\right) \leftrightarrow \alpha \dot{g}\left( t\right) +\beta \dot{f}%
\left( t\right) =\frac{\alpha }{3}\dot{u}\left( t\right);\,\,\left[ \alpha
g\left( t\right) +\beta f\left( t\right) \right] ^{2}=\beta ^{2}u\left(
t\right) ^{2}\leftrightarrow \beta =\frac{\alpha }{3},
\label{alpha-beta}
\end{equation}%
and%
\begin{equation}
\alpha g\left( t\right) =\beta \left[ u\left( t\right) -f\left( t\right) %
\right] \Longleftrightarrow U\left( t\right) =u\left( t\right) \exp \left(
\beta \int u\left( t\right) dt\right) ,  \label{alpha-G}
\end{equation}%
which is indeed the nonlocal transformation in  Eq. (\ref{PT1}) as in  \cite{Chandrasekar 2012,Chandrasekar-PRE 2005}. Yet, $\beta =\omega \,\eta $ not only retrieves the MEE in  Eq. (\ref{q-MEE1}) (readily reported in \cite{Chandrasekar 2012,Chandrasekar-PRE 2005}), but also it retrieves the more general solution 
\begin{equation}
U\left( t\right) =A\sin \left( \omega t+\varphi \right) +B\cos \left( \omega
t+\varphi \right)   \label{HO-gen-solution}
\end{equation}%
for the linear harmonic oscillator dynamical equation in Eq. (\ref{HO-solution}) (i.e., no initial boundary conditions, like $U\left( 0\right)
=0$ as in \cite{Chandrasekar 2006,Chandrasekar 2012,Chandrasekar-PRE 2005}, on Eq. (\ref{HO-solution}) are imposed). Such general solution Eq. (\ref{HO-gen-solution}) would, in a straightforward manner, yield  
\begin{equation}
u\left( t\right) =\frac{A\cos \left( \omega t+\varphi \right) +B\sin \left(
\omega t+\varphi \right) }{1+A\eta \sin \left( \omega t+\varphi \right)
-B\eta \cos \left( \omega t+\varphi \right) }\,;\,\,0\leq \left(
A,B\right) <\frac{1}{\eta },  \label{q-MEE-gen1}
\end{equation}%
which is the most general solutions for Eq.(\ref{q-MEE1}). Moreover, the corresponding canonical momentum is given by Eq.(\ref{q-momentum}).  

In Figure 1(a) we show $u(t)$ of  Eq. (\ref{q-MEE-gen1}) as it evolves in time for different values of the damping parameter $\eta$, where $\omega, A$, and $B$ are fixed. An obvious isochronic oscillatory trend is observed. In Figure 1(b) we show the corresponding phase-space trajectories (using Eq. (\ref{q-MEE-gen1}) and  Eq. (\ref{q-momentum})) for different values of $A$ , where $B=1,\omega=1$, and $\eta=0.15$. The effect of the inherited amplitude $A$ (i.e., inherited from the standard harmonic oscillator general solution in Eq. (\ref{HO-gen-solution})) exhibits the regular phase-space trajectories trend for the MEE Eq. (\ref{q-MEE1}). Similar effect is also observed for different values of the inherited amplitude $B$ for a fixed value of $A$. In figure 1(c), however, we show the effect of the damping related parameter $\eta$ on the phase-space trajectories as they evolve in time.

In the forthcoming sections we shall introduce and report on the PDM counterpart of MEE Eq. (\ref{q-MEE1}). We shall, therefore, leave the initial conditions to be identified by the nature of the corresponding PDM-MEE problem at hand. 

\section{PDM-Modified Emden Equation.}

In this section we discuss the MEE Eq. (\ref{q-MEE}) within position-dependent mass settings. In so doing, we may assume that the coordinate $u$ is deformed in such a way that%
\begin{equation}
u\longrightarrow u\left( x\right) =\int \sqrt{G\left( x\right) }dx=\sqrt{%
F\left( x\right) }x\Longleftrightarrow \sqrt{G\left( x\right) }=\sqrt{%
F\left( x\right) }\left( 1+\frac{F^{\prime }\left( x\right) }{2F\left(
x\right) }x\right) .  \label{1D-PT}
\end{equation}%
One should notice that $u=\sqrt{F(x)}\,x$ suggests that $F(x)$ and $G(x)$ are both positive valued dimensionless scalar multipliers. This would, in turn, imply that $-\infty\leq (u,x) \leq \infty$. Under such deformation/transformation, therefore, one would find that%
\begin{equation}
\dot{u}\longrightarrow \dot{u}\left( x\right) =\sqrt{F\left( x\right) }%
\left( 1+\frac{F^{\prime }\left( x\right) }{2F\left( x\right) }x\right) \dot{%
x}=\sqrt{G\left( x\right) }\dot{x}.  \label{Vq}
\end{equation}%
Which when substituted in Eq. (\ref{q-Lagrangian1}) and Eq. (\ref{q-MEE}), respectively, yields%
\begin{equation}
L\left( u,\dot{u},t\right) \longrightarrow L\left( x,\dot{x},t\right) =\frac{%
1}{2}\,G\left( x\right) \dot{x}^{2}-\left[ \frac{1}{2}\omega ^{2}F\left(
x\right) x^{2}+\frac{1}{4}\beta ^{2}F\left( x\right) ^{2}x^{4}\right] 
\label{PDM-L}
\end{equation}%
and%
\begin{equation}
\ddot{x}+\frac{G^{\prime }\left( x\right) }{2G\left( x\right) }\dot{x}%
^{2}+\alpha \sqrt{F\left( x\right) }x\,\dot{x}+\omega ^{2}x\sqrt{\frac{%
F\left( x\right) }{G\left( x\right) }}\left( 1+\frac{\beta ^{2}}{\omega ^{2}}%
F\left( x\right) x^{2}\right) =0.  \label{PDM-MEE}
\end{equation}%
It should be noted here that the PDM-MEE Eq. (\ref{PDM-MEE}) is a dynamical equation that describes the motion of a PDM-particle moving under the influence of a conservative potential force field $V\left( x\right) =\frac{1%
}{2}\omega ^{2}F\left( x\right) x^{2}+\frac{1}{4}\beta ^{2}F\left( x\right)
^{2}x^{4}$ and subjected to a non-conservative Rayleigh dissipation force
field $\mathcal{R}\left( x,\dot{x}\right) =\frac{1}{2}\alpha \sqrt{F\left(x\right) }\,G\left( x\right) x\,\dot{x}^{2}$. Moreover, equation (\ref{PDM-MEE}) resembles the well known mixed Li\'{e}nard-type differential equation%
\begin{equation}
\ddot{x}+a\left( x\right) \dot{x}^{2}+b\left( x\right) \dot{x}+c\left(
x\right) =0,  \label{PDM mixed-Lienard}
\end{equation}%
with,%
\begin{equation*}
\,a\left( x\right) =\frac{G^{\prime }\left( x\right) }{2G\left( x\right) }%
,\,\,b\left( x\right) =\alpha \sqrt{F\left( x\right) }x,\,\, c\left( x\right)
=\omega ^{2}x\sqrt{\frac{F\left( x\right) }{G\left( x\right) }}\left( 1+%
\frac{\beta ^{2}}{\omega ^{2}}F\left( x\right) x^{2}\right) .
\end{equation*}%
Hence, the transition from a mixed Li\'{e}nard-type differential equation (\ref{PDM mixed-Lienard}) into a MEE type Eq. (\ref{q-MEE}) (the other way round is also viable) is feasible through the transformation Eq. (\ref{1D-PT}). Moreover, the PDM quasi-MEE Eq. (\ref{PDM-MEE}) consequently inherits solution Eq. (\ref{q-MEE-solution}) of MEE Eq. (\ref{q-MEE}) through the point transformation Eq. (\ref{1D-PT}). That is, once the relation between $u\left(
t\right) $ and $x\left( t\right) $ is identified, for a given PDM particle $G\left( x\right) =G\left( x\left( t\right) \right) $, then the structure of Eq. (\ref{PDM-MEE}) collapses into that of Eq. (\ref{q-MEE}) and consequently the exact solution of the PDM-MEE Eq. (\ref{PDM-MEE}) would be obtained in the process, therefore. We illustrate this issue in the following examples.

\subsection{MEE-type Mathews-Lakshmanan PDM $G\left( x\right) =1/\left( 1+\lambda^{2}x^{2}\right) $}

A Mathews-Lakshmanan \cite{M-L 1974} PDM particle, with a positive valued dimensionless scalar multiplier $G\left( x\right) =1/\left( 1+\lambda ^{2}x^{2}\right) $, would lead, through $F\left( x\right) $ and $G\left( x\right) $ correlation Eq. (\ref{1D-PT}), to%
\begin{equation}
u=\sqrt{F(x)}\,x\Rightarrow\sqrt{F\left( x\right) }x=\frac{1}{\lambda }\ln \left( \lambda x+\sqrt{%
1+\lambda ^{2}x^{2}}\right).
\end{equation}%
At this point one should notice that $\lambda$ is a coupling parameter that renders the scalar multipliers $F(x)$ and $G(x)$ dimensionless and positive-valued (i.e., $\lambda$ can never be zero, otherwise the dynamical system will collapse).  The corresponding PDM quasi -MEE dynamical equation (\ref{PDM-MEE}) reads%
\begin{equation}
\ddot{x}-\frac{\lambda ^{2}x}{1+\lambda ^{2}x^{2}}\dot{x}^{2}+\frac{3\eta
\omega }{\lambda }\,\mathcal{Z}\left( x\right) \,\dot{x}+\frac{\omega ^{2}}{%
\lambda }\sqrt{1+\lambda ^{2}x^{2}}\mathcal{Z}\left( x\right) \left[ 1+\frac{%
\eta ^{2}}{\lambda ^{2}}\mathcal{Z}\left( x\right) ^{2}\right] =0,
\label{PDM-ML-MEE}
\end{equation}%
where,%
\begin{equation*}
\mathcal{Z}\left( x\right) =\ln \left( \lambda x+\sqrt{1+\lambda ^{2}x^{2}}%
\right),
\end{equation*}%
is used. Hence, 
\begin{equation}
u\left( t\right) =\sqrt{F\left( x\right) }x=\frac{\mathcal{Z}\left( x\right) 
}{\lambda }\Longleftrightarrow x\left( t\right) =\frac{1}{\lambda }\sinh
\left( \lambda u\left( t\right) \right) ,  \label{ML-x(t)}
\end{equation}%
where $u\left( t\right) $ is given by Eq. (\ref{q-MEE-gen1}). Hereby, one should notice that $\lambda$ is a coupling parameter that renders the positive-valued scalar multipliers $G(x)$ and $F(x)$ dimensionless. 
Nevertheless, one should be advised that the substitution of the solution $x\left( t\right) $ of Eq.\ref{ML-x(t)}) would transform Eq. (\ref{PDM-ML-MEE}) into the MEE  Eq. (\ref{q-MEE1}) (with its solution readily given in Eq. (\ref{q-MEE-solution})). The exact solvability of Eq. (\ref{PDM-ML-MEE}) is secured, therefore.  

Moreover,  the dynamical equation (\ref{PDM-ML-MEE}) represents a Mathews-Lakshmanan PDM-particle, $G\left( x\right) =1/\left( 1+\lambda ^{2}x^{2}\right) $, moving under the influence of a conservative PDM-deformed potential force field%
\begin{equation}
V\left( x\right) =\frac{1}{2}\omega
^{2}\left( \frac{1}{\lambda }\ln \left( \lambda x+\sqrt{1+\lambda ^{2}x^{2}}%
\right) \right) ^{2}
\label{V-ML}
\end{equation}%
and a non-conservative PDM-deformed Rayleigh dissipative force field%
\begin{equation}
\mathcal{R}\left( x,\dot{x}\right) =\frac{3\eta
\omega }{2\lambda }\, \frac{\mathcal{Z}\left( x\right)}{\left( 1+\lambda ^{2}x^{2}\right)}\dot{x}^{2}.
\label{R-ML}
\end{equation}
The standard PDM Lagrangian and PDM-Hamiltonian for this system, respectively read,%
\begin{equation}
L=\frac{1}{2}\frac{\dot{x}^{2}}{1+\lambda ^{2}x^{2}}-\frac{1}{2}\omega
^{2}\left( \frac{1}{\lambda }\ln \left( \lambda x+\sqrt{1+\lambda ^{2}x^{2}}%
\right) \right) ^{2},  \label{PDM-ML-L/H}
\end{equation}%
\begin{equation*}
H=\frac{p^{2}}{2}\left( 1+\lambda ^{2}x^{2}\right) +\frac{1}{2}\omega
^{2}\left( \frac{1}{\lambda }\ln \left( \lambda x+\sqrt{1+\lambda ^{2}x^{2}}%
\right) \right) ^{2},
\end{equation*}%
where the PDM canonical momentum in this case reads%
\begin{equation}
p\left( t\right) =\frac{\dot{x}\left( t\right) }{1+\lambda ^{2}x\left(
t\right) ^{2}}=\frac{\dot{u}\left( t\right) }{\cosh \left( \lambda u\left(
t\right) \right) }.  \label{PDM-ML-momentum}
\end{equation}%
In Figure 2, we plot $x(t)$ of Eq. (\ref{ML-x(t)}) as it evolves in time with  $\omega=2$ and $\varphi=0$,   for different values of $\eta$ at $A=B=\lambda=1$ in Fig. 2(a), for different values of $\lambda$ at $\eta=0.5$ and $A=B=1$ in Fig. 2(b), and for different values of $B$ at $\eta=0.25$, $\lambda=2$, and $A=1$ in Fig. 2(c). We observe that the oscillations are still isochronic, documenting again that an amplitude -dependent frequency is not a consequential property of the nonlinear dynamics even under PDM-settings. We, moreover, show the corresponding phase-space trajectories for different values of $\eta$ at $A=1$, $B=0.25$, and $\lambda=2$ in Fig. 2(d),  for different values of $\lambda$ at $\eta=0.5$ and $A=B=0.25$ in Fig. 2(e), and for different values of $B$ at $\eta=0.05$, $\lambda=2$, and $A=1$ in Fig. 2(f).

\subsection{An exponential MEE-type PDM $G\left( x\right) =e^{2\lambda x}$}

A PDM with a positive valued dimensionless scalar multiplier in the form of $G\left( x\right) =e^{2\lambda x}$, where $\lambda$ is as defined above, would lead to%
\begin{equation}
F\left( x\right) =\frac{e^{2\lambda x}}{\lambda ^{2}x^{2}}\left(
1-e^{-\lambda x}\right) ^{2},  \label{Q-exponential}
\end{equation}%
and results%
\begin{equation}
x\left( t\right) =\frac{1}{\lambda }\ln \left( \lambda u\left( t\right)
+1\right) \,,  \label{x-exponential}
\end{equation}%
where $u\left( t\right) $ is given by Eq. (\ref{q-MEE-solution}). The substitution of $x\left( t\right) $ of Eq. (\ref{x-exponential}) in the corresponding Euler-Lagrange equation%
\begin{equation}
\ddot{x}+\lambda \dot{x}^{2}+\frac{3\eta \omega }{\lambda }e^{\lambda
x}\left( 1-e^{-\lambda x}\right) \,\dot{x}+\frac{\omega _{\circ }^{2}}{%
\lambda }\left( 1-e^{-\lambda x}\right) \left[ 1+\frac{\eta ^{2}}{\lambda
^{2}}e^{2\lambda x}\left( 1-e^{-\lambda x}\right) ^{2}\right] =0,
\label{EL-exponential}
\end{equation}%
implies the dynamical equation (\ref{PDM-MEE}) with its readily existing exact solution in Eq. (\ref{q-MEE-solution}). The dynamical equation (\ref{EL-exponential}) represents a PDM-particle, $G\left( x\right) =e^{2\lambda x}$, moving under the influence of a conservative PDM-deformed potential force field%
\begin{equation}
V\left( x\right) =\frac{1}{2}\omega _{\circ }^{2}\,\frac{e^{2\lambda x}}{%
\lambda ^{2}}\left( 1-e^{-\lambda x}\right) ^{2}+\frac{1}{4}\beta ^{2}\,\frac{%
e^{4\lambda x}}{\lambda ^{4}}\left( 1-e^{-\lambda x}\right) ^{4},
\label{V-exponential}
\end{equation}%
and a non-conservative PDM-deformed Rayleigh dissipative force field%
\begin{equation}
\mathcal{R}\left( x,\dot{x}\right) =\frac{3\eta \omega }{2\lambda }%
e^{2\lambda x}\left( e^{\lambda x}-1\right) \,\dot{x}^{2}.
\label{R-exponential}
\end{equation}%
Moreover, the standard PDM-Lagrangian and PDM-Hamiltonian that describe this PDM-particle are given, respectively, by%
\begin{equation}
L=\frac{1}{2}e^{2\lambda x}\dot{x}^{2}-\left[ \frac{1}{2}\omega _{\circ }^{2}%
\frac{e^{2\lambda x}}{\lambda ^{2}}\left( 1-e^{-\lambda x}\right) ^{2}+\frac{%
1}{4}\beta ^{2}\frac{e^{4\lambda x}}{\lambda ^{4}}\left( 1-e^{-\lambda
x}\right) ^{4}\right] ,  \label{L-exponential}
\end{equation}%
\begin{equation}
H=\frac{1}{2}e^{-2\lambda x}p\left( x\right) ^{2}+\frac{1}{2}\omega _{\circ
}^{2}\frac{e^{2\lambda x}}{\lambda ^{2}}\left( 1-e^{-\lambda x}\right) ^{2}+%
\frac{1}{4}\beta ^{2}\frac{e^{4\lambda x}}{\lambda ^{4}}\left( 1-e^{-\lambda
x}\right) ^{4},\,  \label{H-ecponential}
\end{equation}%
where%
\begin{equation}
p\left( t\right) =e^{2\lambda x\left( t\right) }\,\dot{x}\left( t\right) ,
\label{p-exponential}
\end{equation}%
is the PDM canonical momentum.

In Figure 3, we plot $x(t)$ of Eq. (\ref{x-exponential}) with  $\omega=1$, and $\varphi=0$   for different values of $\eta$ at $A=B=0.25$, $\lambda=1.5$ in Fig. 3(a), for different values of $\lambda$ at $\eta=0.05$ and $A=B=0.5$ in Fig. 3(b), and for different values of $A$ at $\eta=0.25$, $\lambda=0.5$, and $B=0.5$ in Fig. 3(c). The oscillations are still isochronic even under PDM-settings. The corresponding phase-space trajectories for different values of $\eta$ at $A=B=0.25$, , and $\lambda=1.5$ in Fig.3(d),  for different values of $\lambda$ at $\eta=0.05$ and $A=B=0.5$ in Fig. 3(e), and for different values of $A$ at $\eta=0.05$, $\lambda=1.2$, and $B=0.5$ in Fig. 3(f).

\section{Concluding Remarks}

In this article, we have used the most general solution for the simple harmonic oscillator's dynamical equation (i.e., the initial conditions to be identified by the PDM quasi-MEE problem at hand) and reported the most general solution for the modified Emden equation (\ref{q-MEE1}). This solution is valid as long as the MEE is linearized into the HO dynamical equation (\ref{HO-solution}) through the nonlocal transformation Eq. (\ref{PT2-gen}). Moreover, We have introduced the corresponding PDM quasi--MEE dynamical systems Eq. (\ref{PDM-MEE}) through the transformation Eq. (\ref{1D-PT}). Two illustrative examples are considered and their exact solutions, Eq. (\ref{ML-x(t)}) and Eq. (\ref{x-exponential}), as well as the corresponding phase-space trajectories are reported.%

In connection with the certain unusual nonlinear properties of the MEE, we have observed that the corresponding PDM quasi-MEE continues to exhibit such unusual properties and isochronic oscillations similar to the linear harmonic oscillator. 

On the linearizability side of the MEE Eq. (\ref{q-MEE}), one may start with the damped harmonic oscillator (DHO) dynamical equation%
\begin{equation}
\ddot{U}+2\zeta\dot{U}+\omega ^{2}U=0,  \label{DHO-eqn}
\end{equation}%
where $\zeta=\omega\,\eta$ as in \cite{Mustafa Phys.Scr. 2021} and use a nonlocal transformation (a special case of the ones reported by Chandrasekar et al. in \cite{Chandrasekar 2006}) of the form%
\begin{equation}
U=u\left( t\right) e^{-\zeta \,t} \exp \left( \beta\int 
u\left( t\right) dt\right),  \label{PT3-gen}
\end{equation}%
to obtain%
\begin{equation}
\ddot{u}\left( t\right) +3\beta \,u\left( t\right) \,\dot{u}\left(
t\right) +\Omega ^{2}u\left( t\right) +\beta ^{2}u\left( t\right)
^{3}=0,  \label{q-MEE3}
\end{equation}%
where $\Omega^{2}=\omega^2(1-\eta^2)$ and $\beta=\eta \, \Omega$. This would, in turn, suggest a solution for the new MEE Eq. (\ref{MEE3}) as%
\begin{equation}
u\left( t\right) = \left(  \frac{A\cos \left( \Omega t+\varphi \right) +B\sin \left(
\Omega t+\varphi \right) }{1+A\eta \sin \left( \Omega t+\varphi \right)
-B \eta \cos \left( \Omega t+\varphi \right) } \right)\,;\,\, 0\leq \left(
A,B\right) <\frac{1}{\eta}.  \label{q-MEE3-gen}
\end{equation}%
Obviously, the frequency of oscillation $\Omega$ is now $\eta$-dependent (i.e., frictional related parameter) and consequently would yield non-isochronic oscillations as it evolve in time. 

Finally, such a nonlinear PDM quasi-MME Eq. (\ref{PDM-MEE})  may find its applicability for more complicated nonlinear dynamical systems, for example, in the equilibrium configuration of spherical gas clouds \cite{Leach 1985}, in the spherically symmetric expansion or collapse of relativistically gravitating mass \cite{Mc Vittie 1980}, in the one-dimensional analog of the Yang-Mills boson gauge theory \cite{Yang-Mills 1954}, etc. To the best of our knowledge, this work has never been discussed elsewhere.

\begin{figure}[h!]  
\centering
\includegraphics[width=0.45\textwidth]{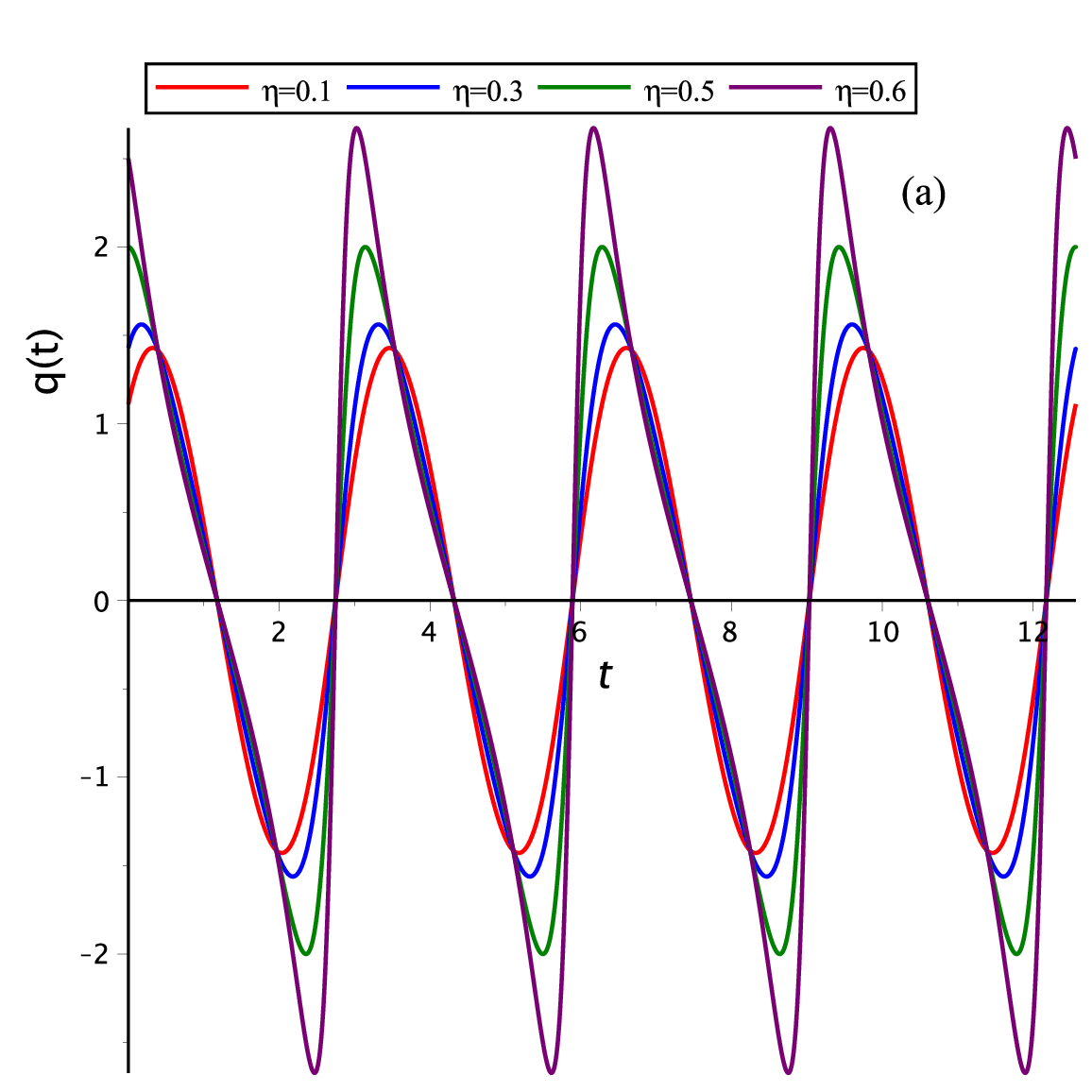}
\includegraphics[width=0.45\textwidth]{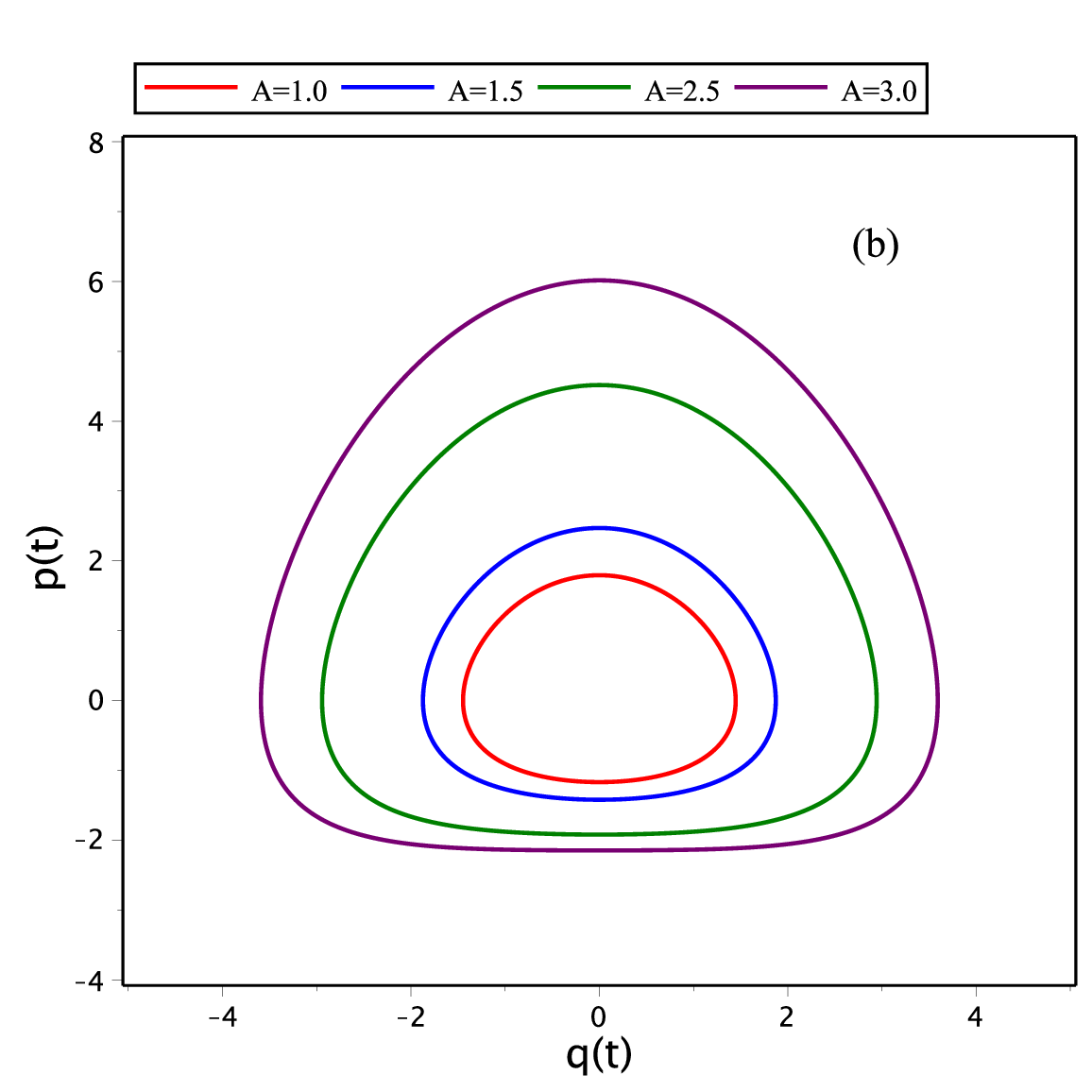} 
\includegraphics[width=0.45\textwidth]{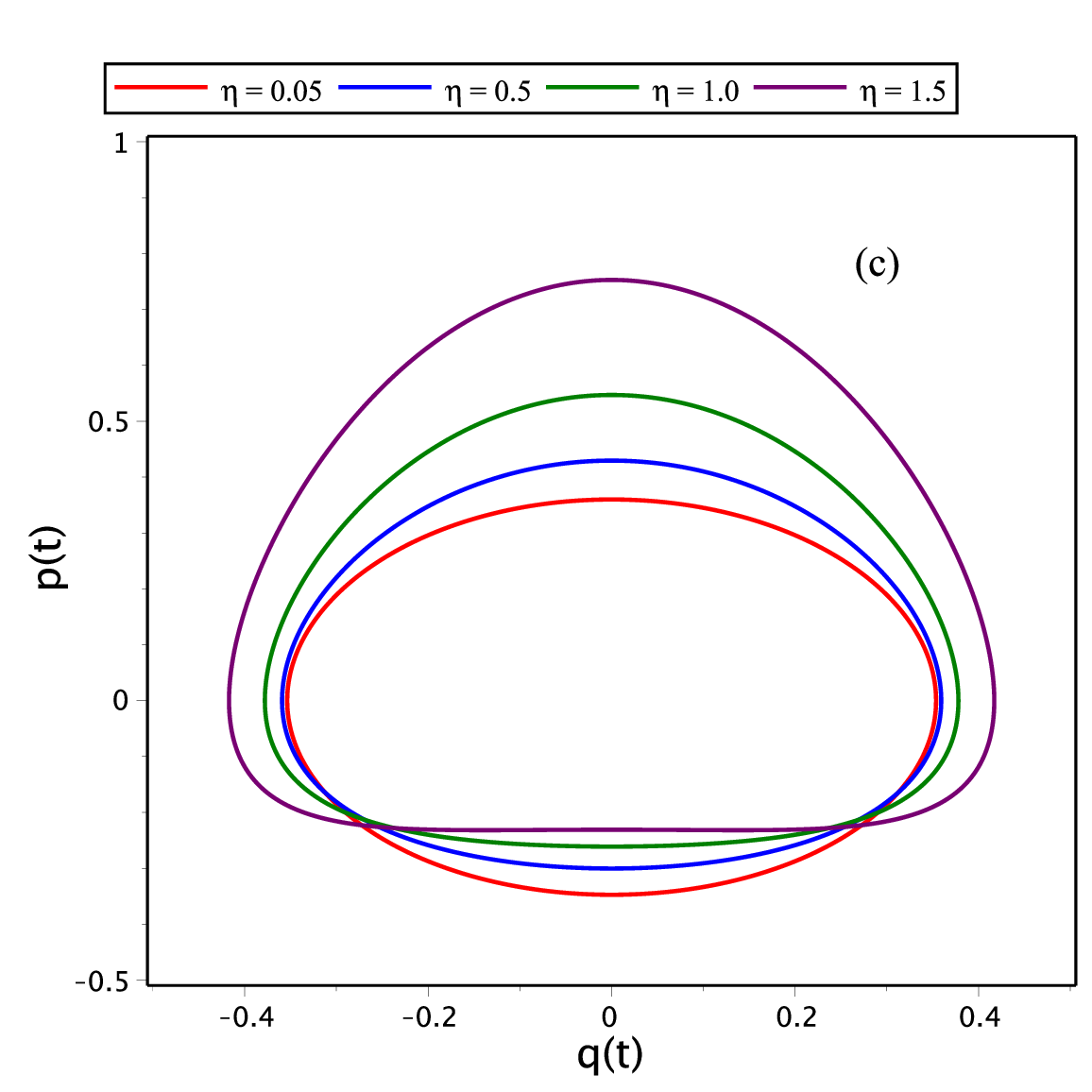}
\caption{ \small
{With $\varphi=0$, we show (a) $q(t)\equiv u(t)$ of  (\ref{q-MEE-gen1}) for different values of $\eta$ with $\omega=2, A=B=1$. (b) The corresponding phase-space trajectory for different values of $A$ with $B=1,\omega=1$, and $\eta=0.15$, and (c) the phase-trajectory for different values of $\eta$ with $\omega=1$, and $A=B=0.25$.}}
\label{fig1}
\end{figure}%
\begin{figure}[h!]  
\centering
\includegraphics[width=0.45\textwidth]{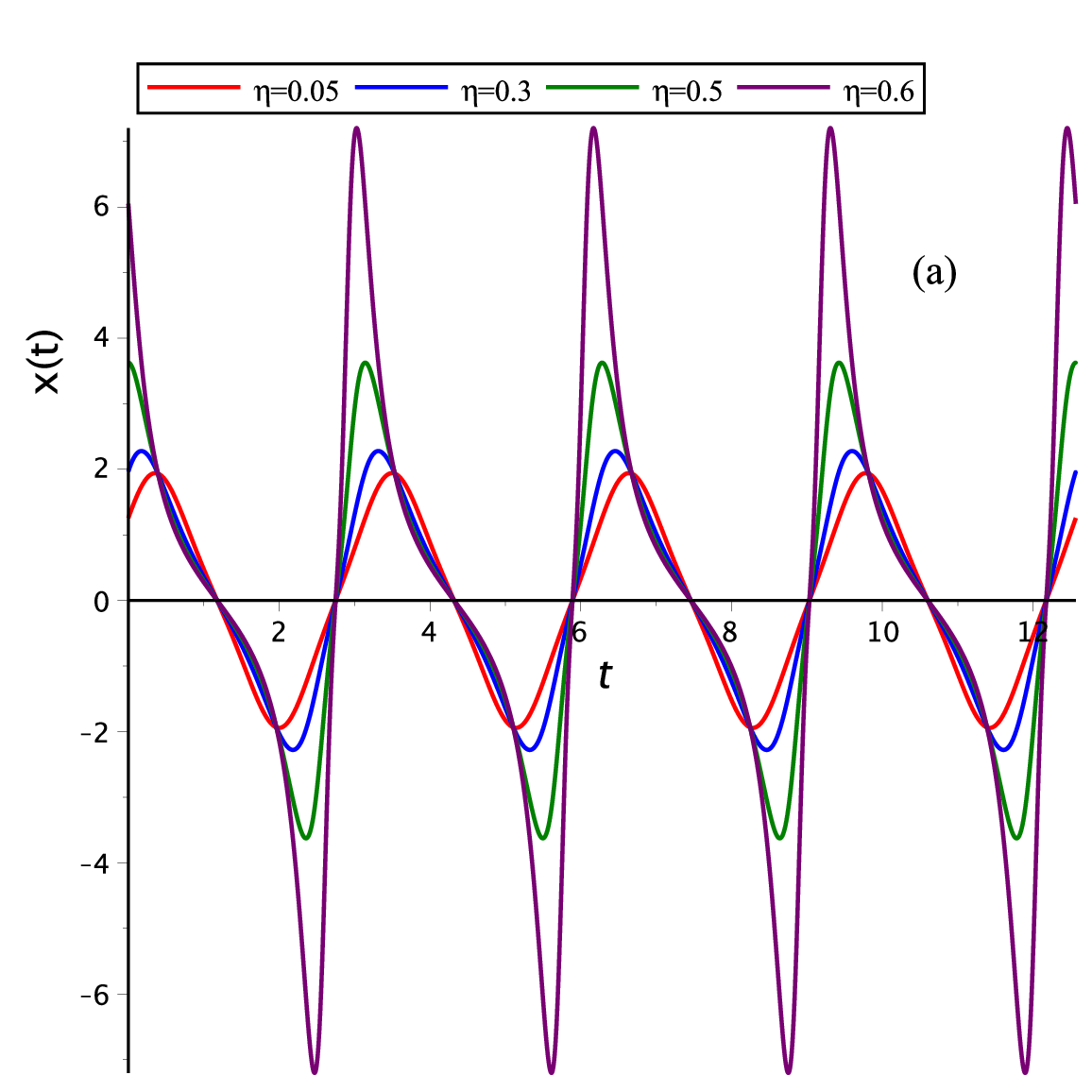}
\includegraphics[width=0.45\textwidth]{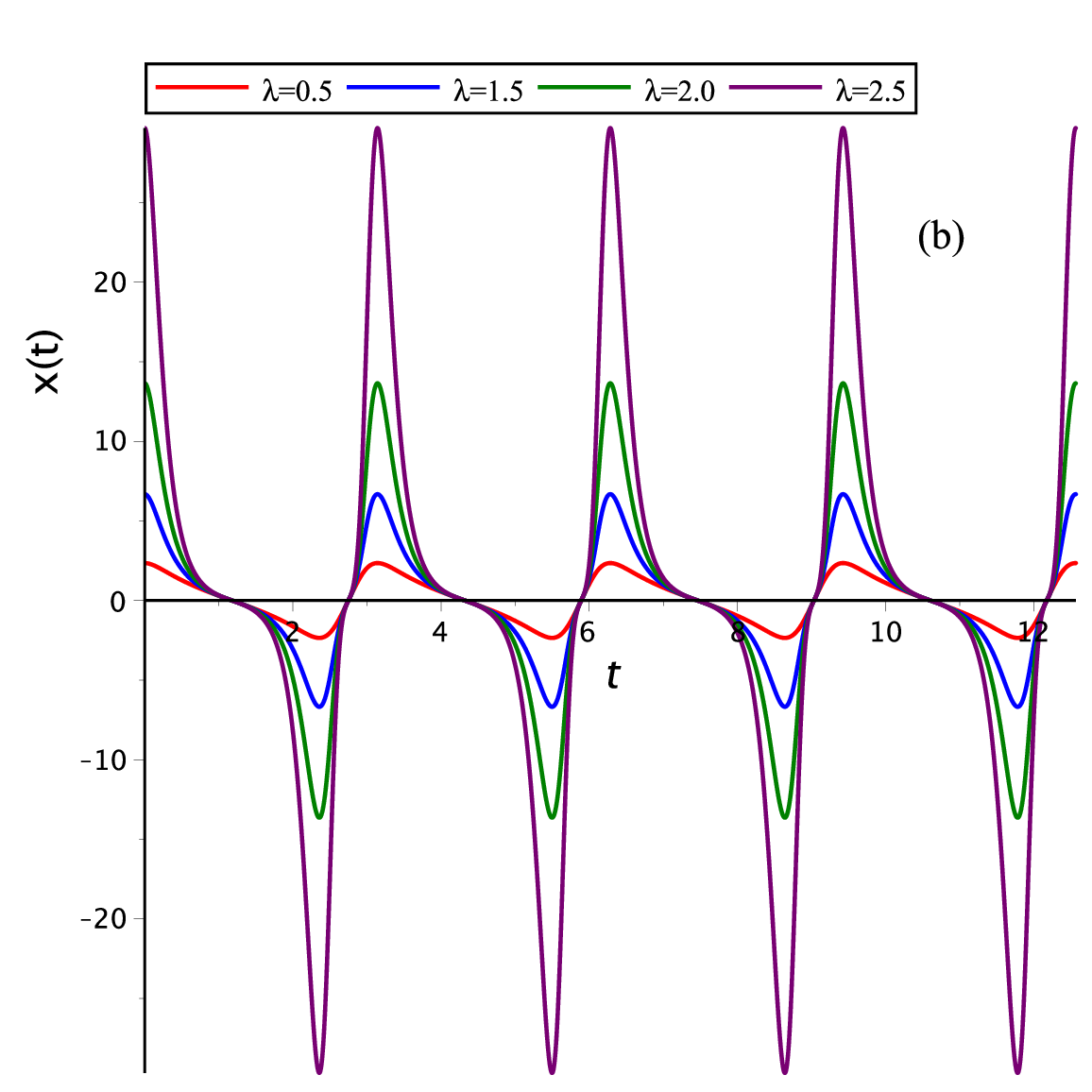} 
\includegraphics[width=0.45\textwidth]{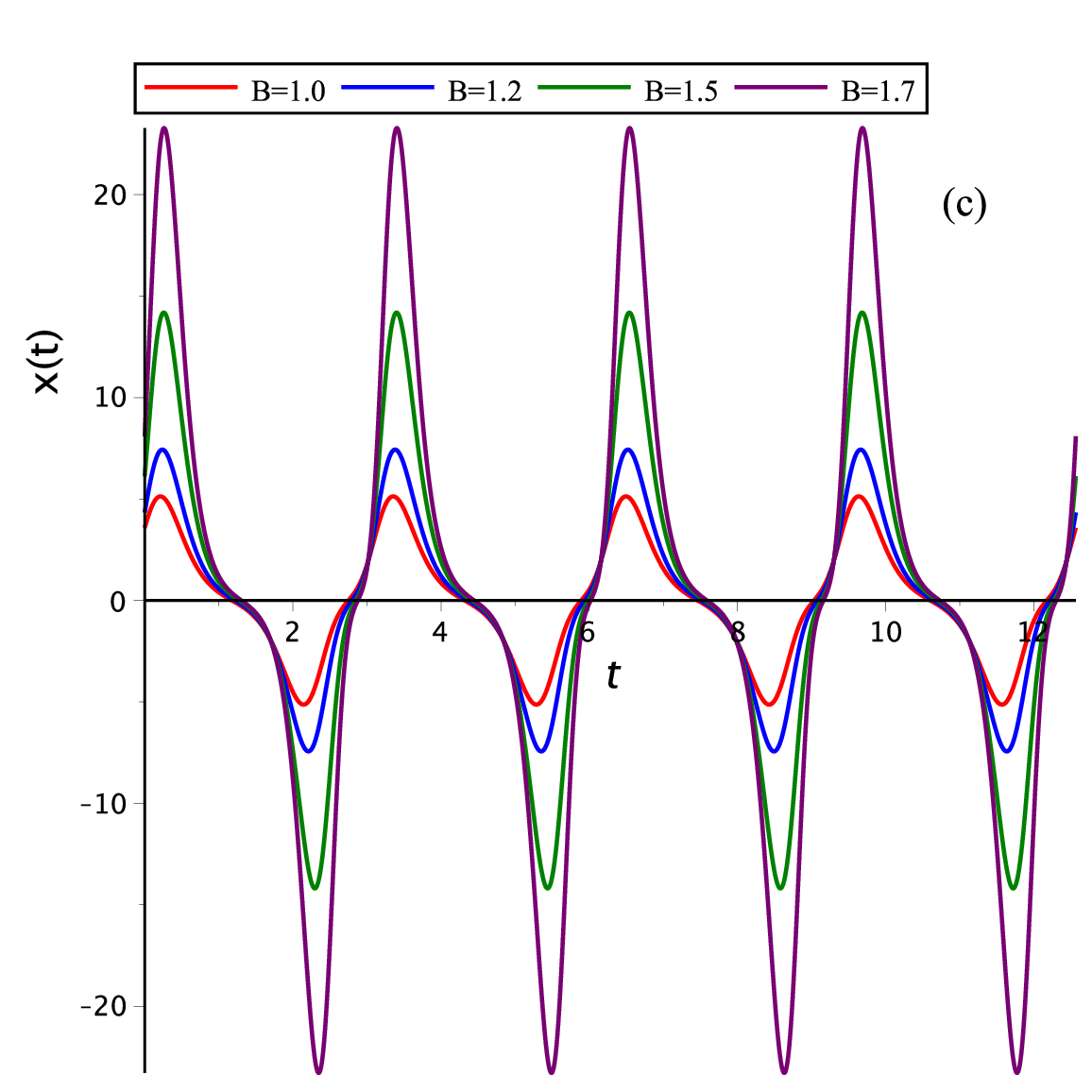}
\includegraphics[width=0.45\textwidth]{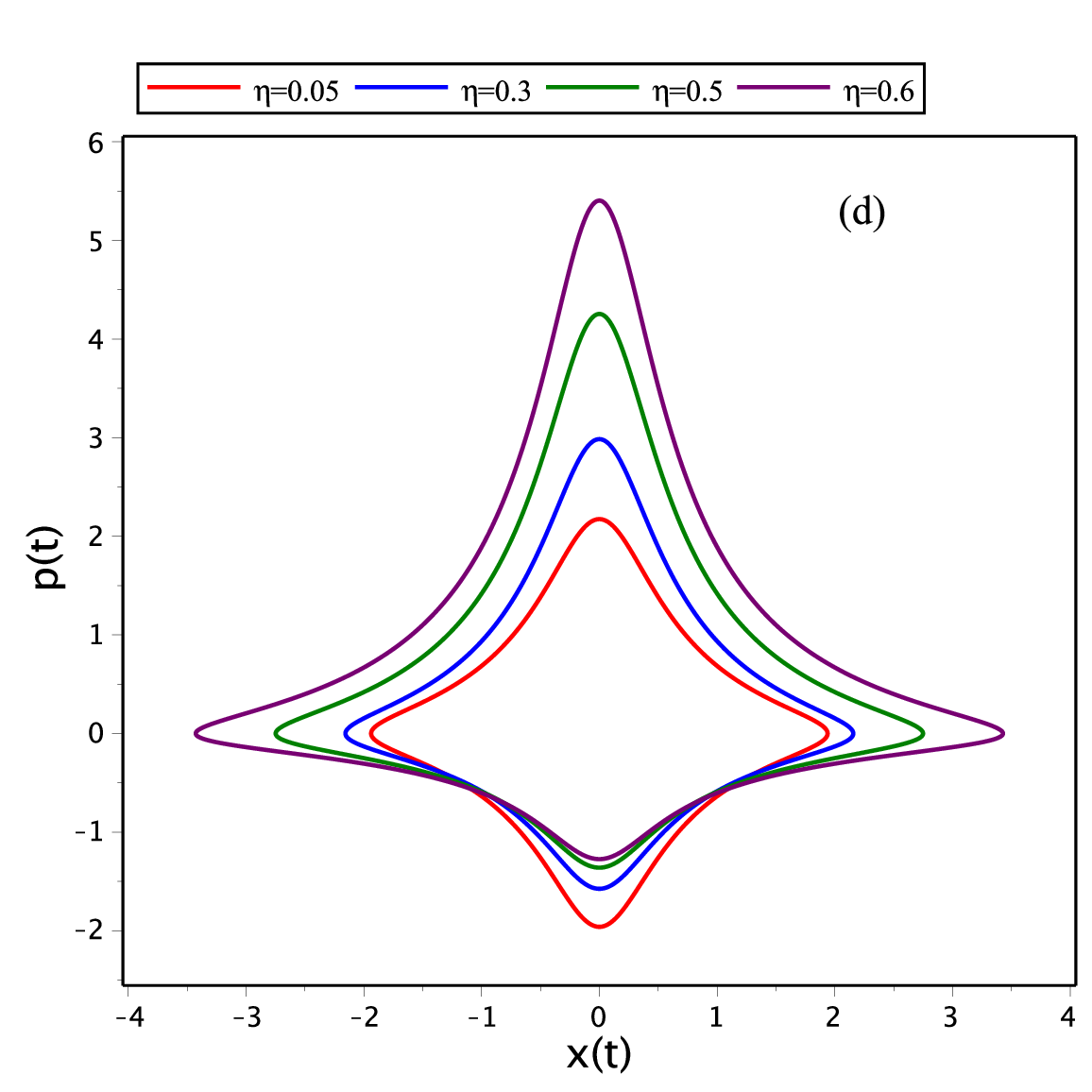}
\includegraphics[width=0.45\textwidth]{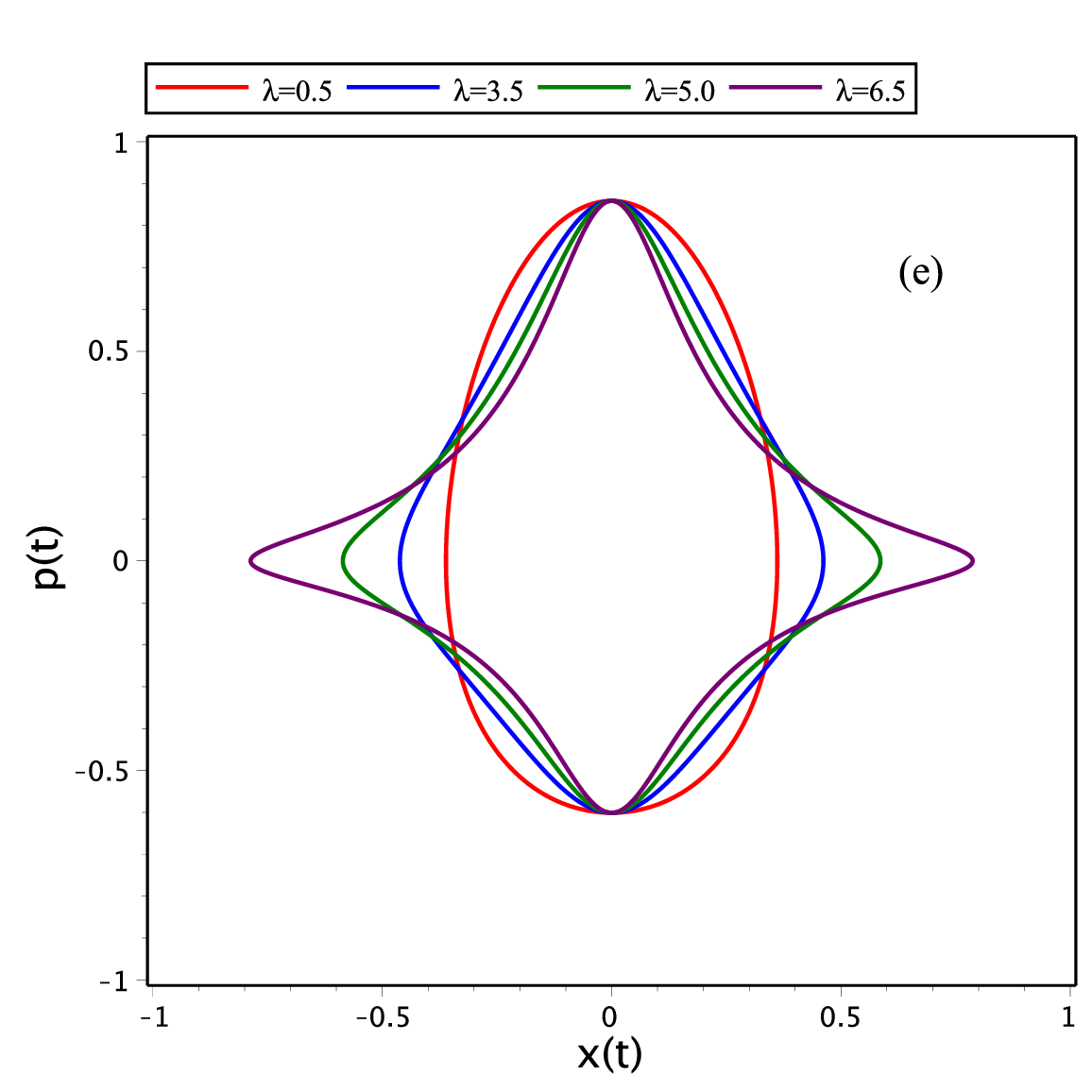} 
\includegraphics[width=0.45\textwidth]{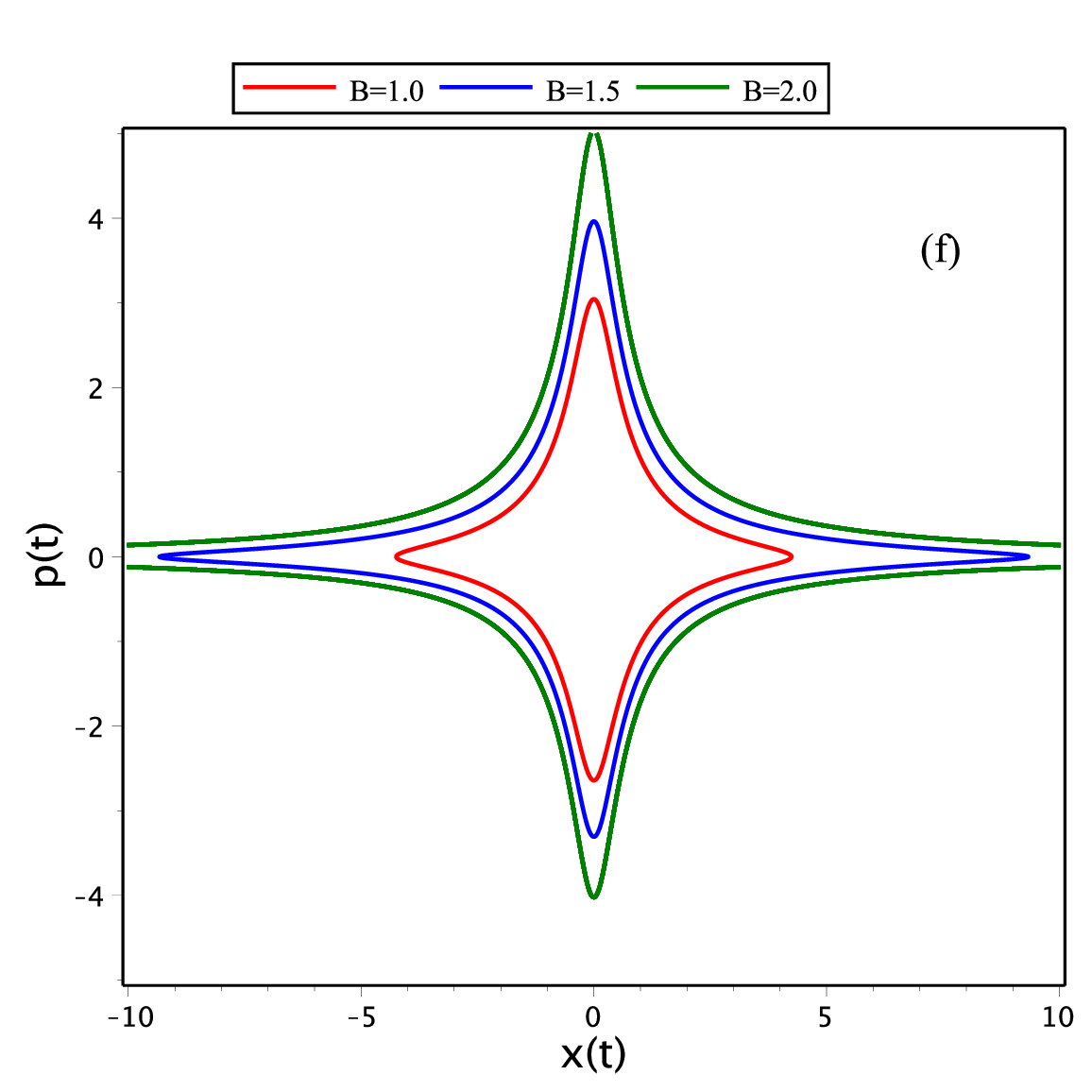}
\caption{\small 
{With $\varphi=0$, we show (a) $x(t)$ of (\ref{ML-x(t)}) for different values $\eta$, with $\omega=2$, $A=B=1$, and $\lambda=1$, (b) $x(t)$ of (\ref{ML-x(t)}) for different values $\lambda$, with $\omega=2$, $A=B=1$, and $\eta=0.5$, (c) $x(t)$ of (\ref{ML-x(t)}) for different values $B$, with $\omega=2$, $A=B=1$,  $\eta=0.25$, and $\lambda=2$. The corresponding phase-space trajectories are shown in (d) for different values of $\eta$ with $A=1$, $B=0.25$, $\lambda=\omega=2$, (e) for different values of $\lambda$ with $A=B=0.25$,  $\omega=2$, $\eta=0.5$, and (f) for different values of $B$ with $A=1$, $\eta=0.05$, $\lambda=\omega=2$. }}
\label{fig2}
\end{figure}%
\begin{figure}[h!]  
\centering
\includegraphics[width=0.45\textwidth]{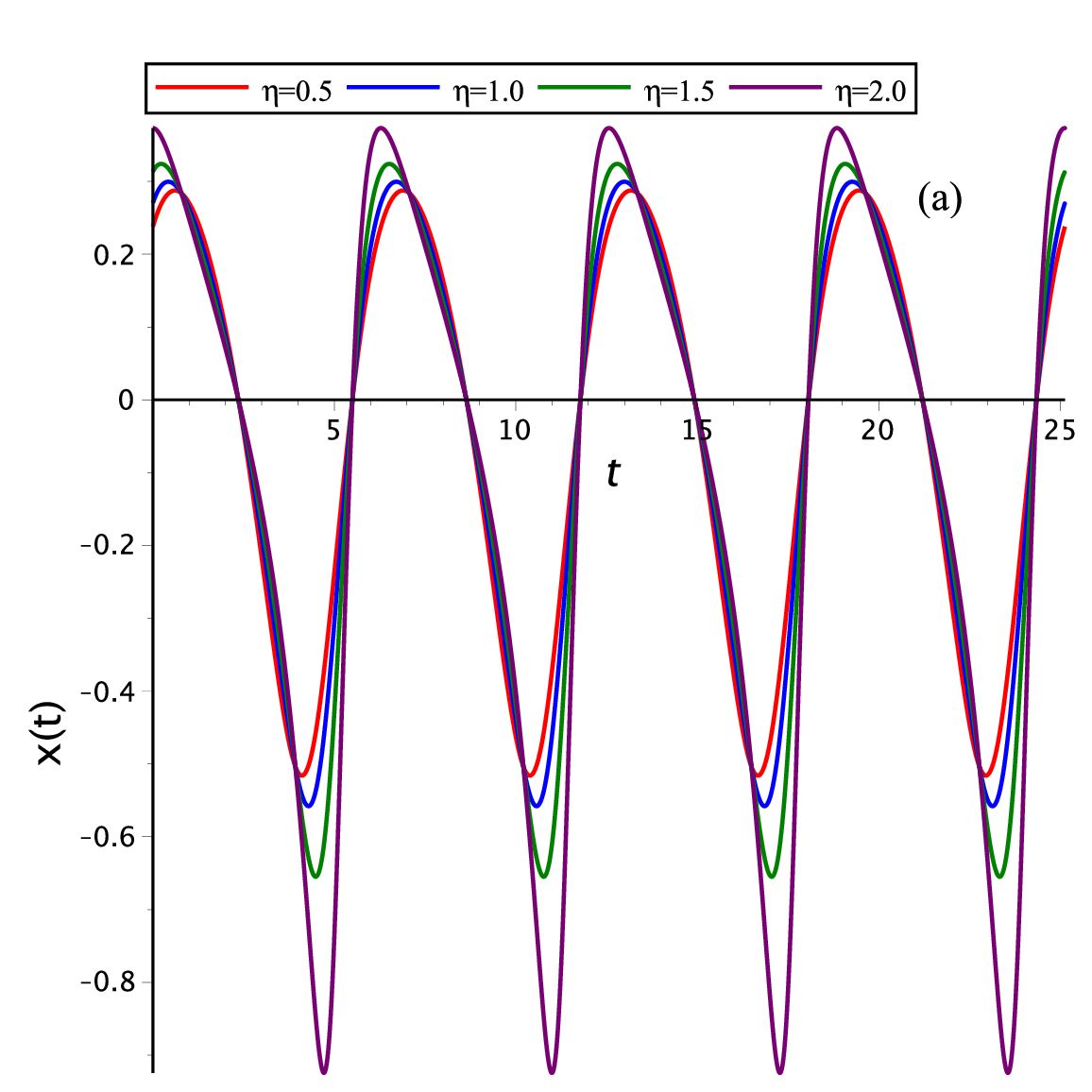}
\includegraphics[width=0.45\textwidth]{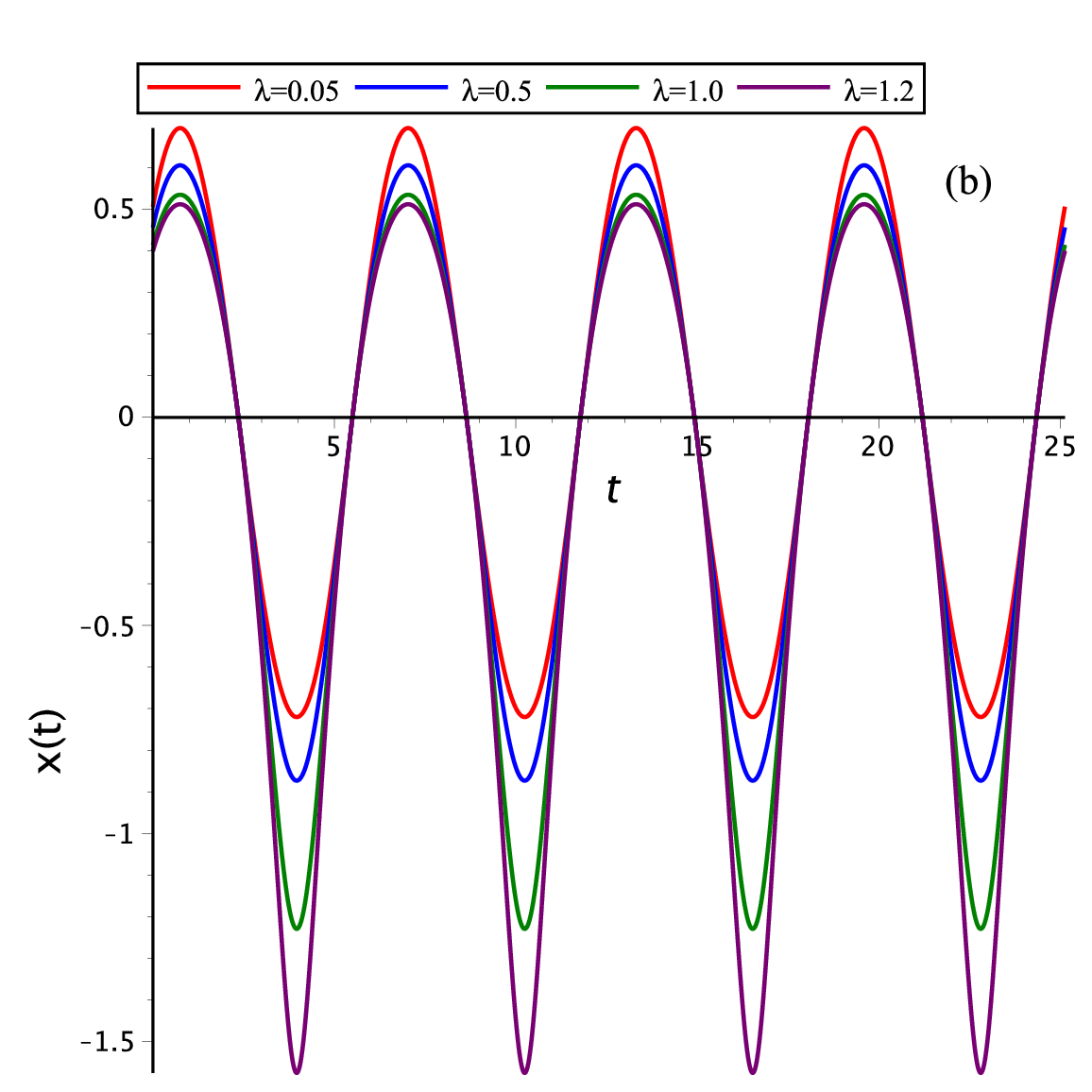} 
\includegraphics[width=0.45\textwidth]{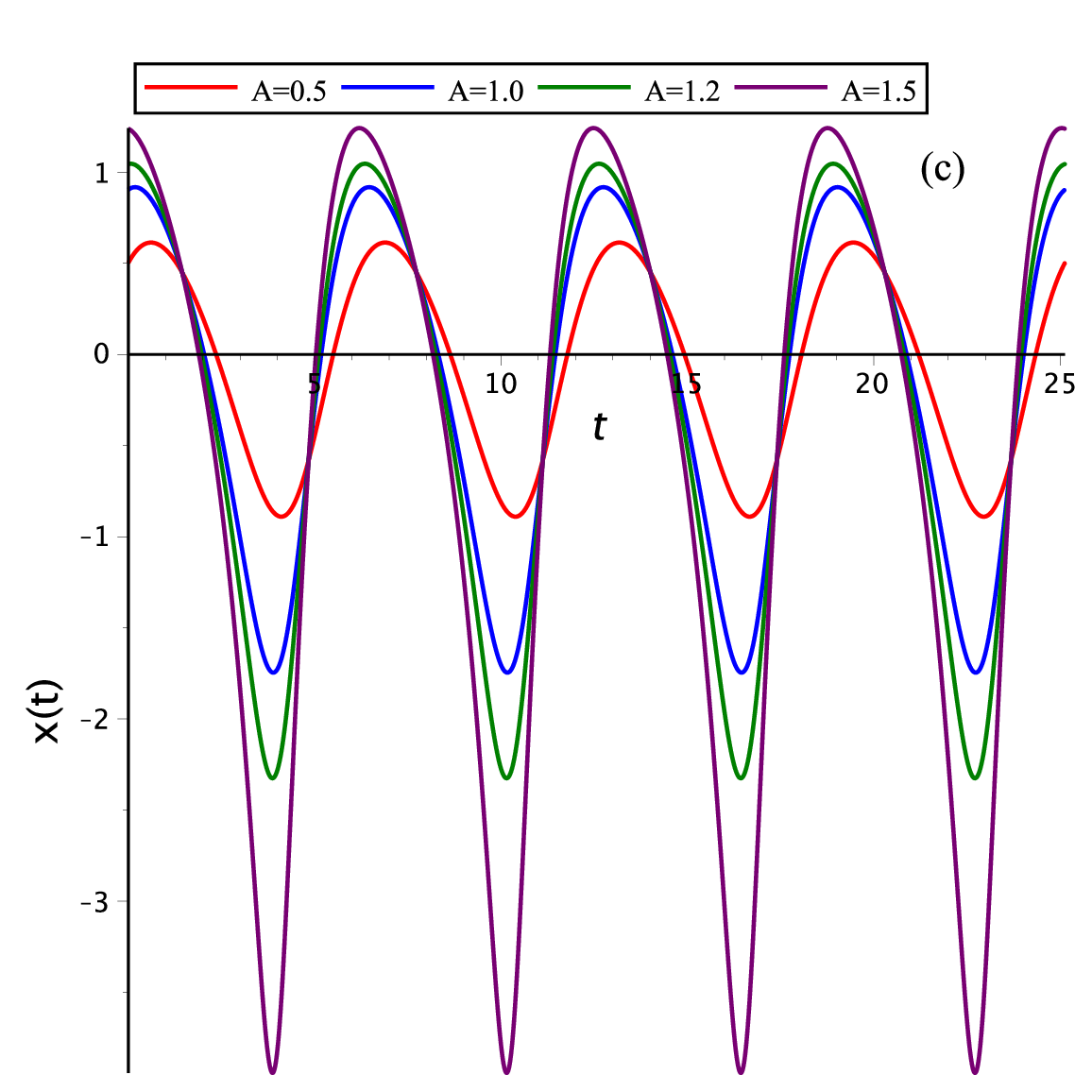}
\includegraphics[width=0.45\textwidth]{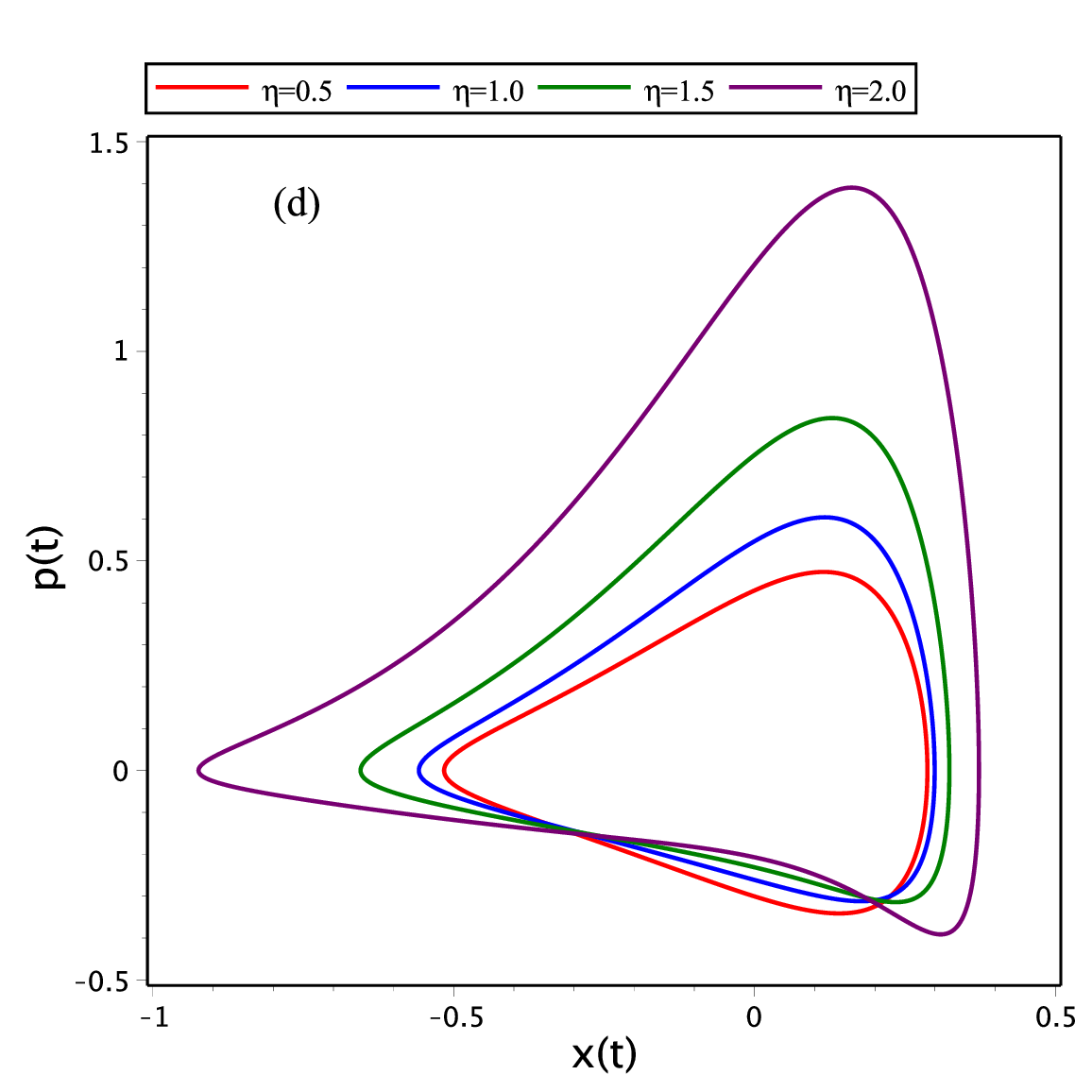}
\includegraphics[width=0.45\textwidth]{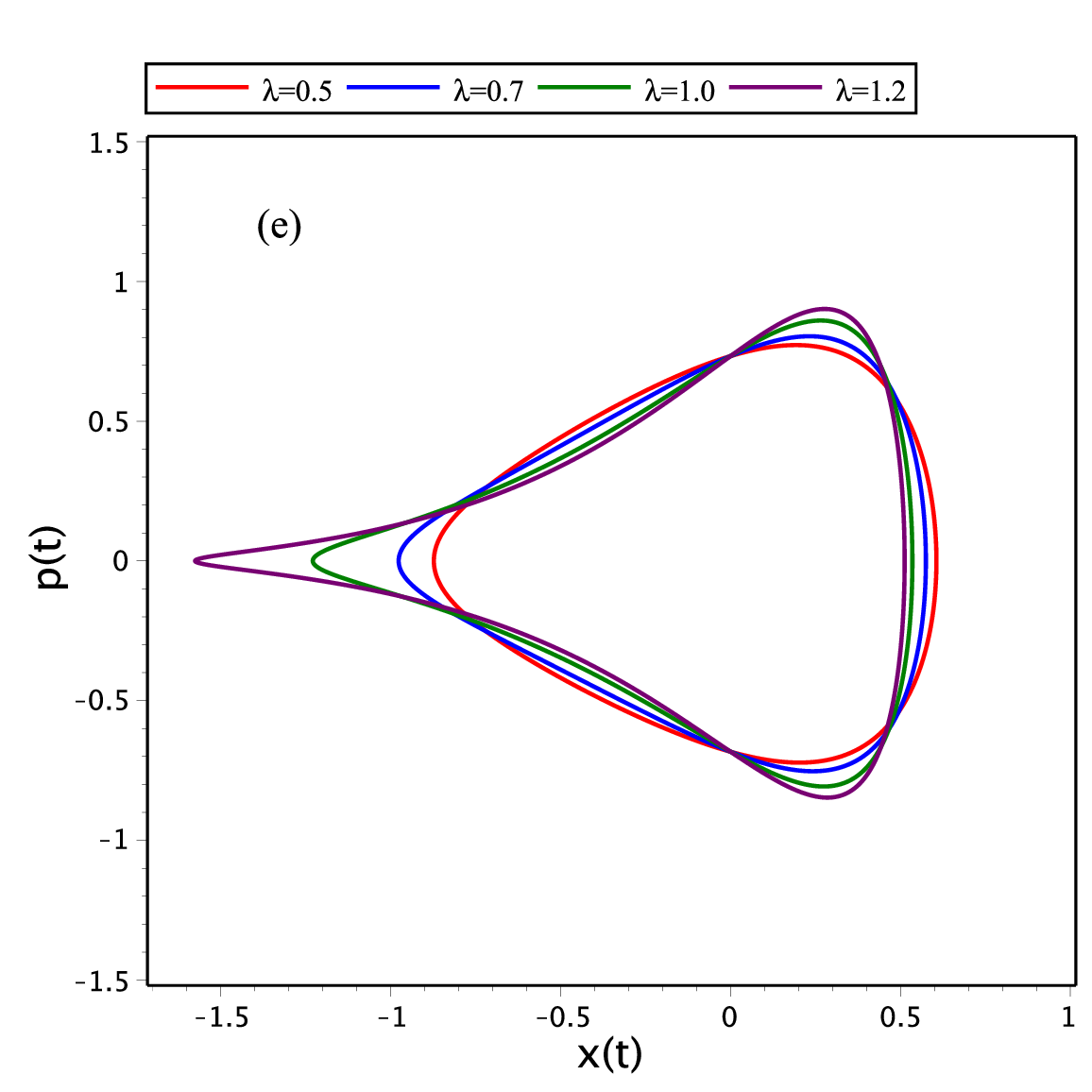} 
\includegraphics[width=0.45\textwidth]{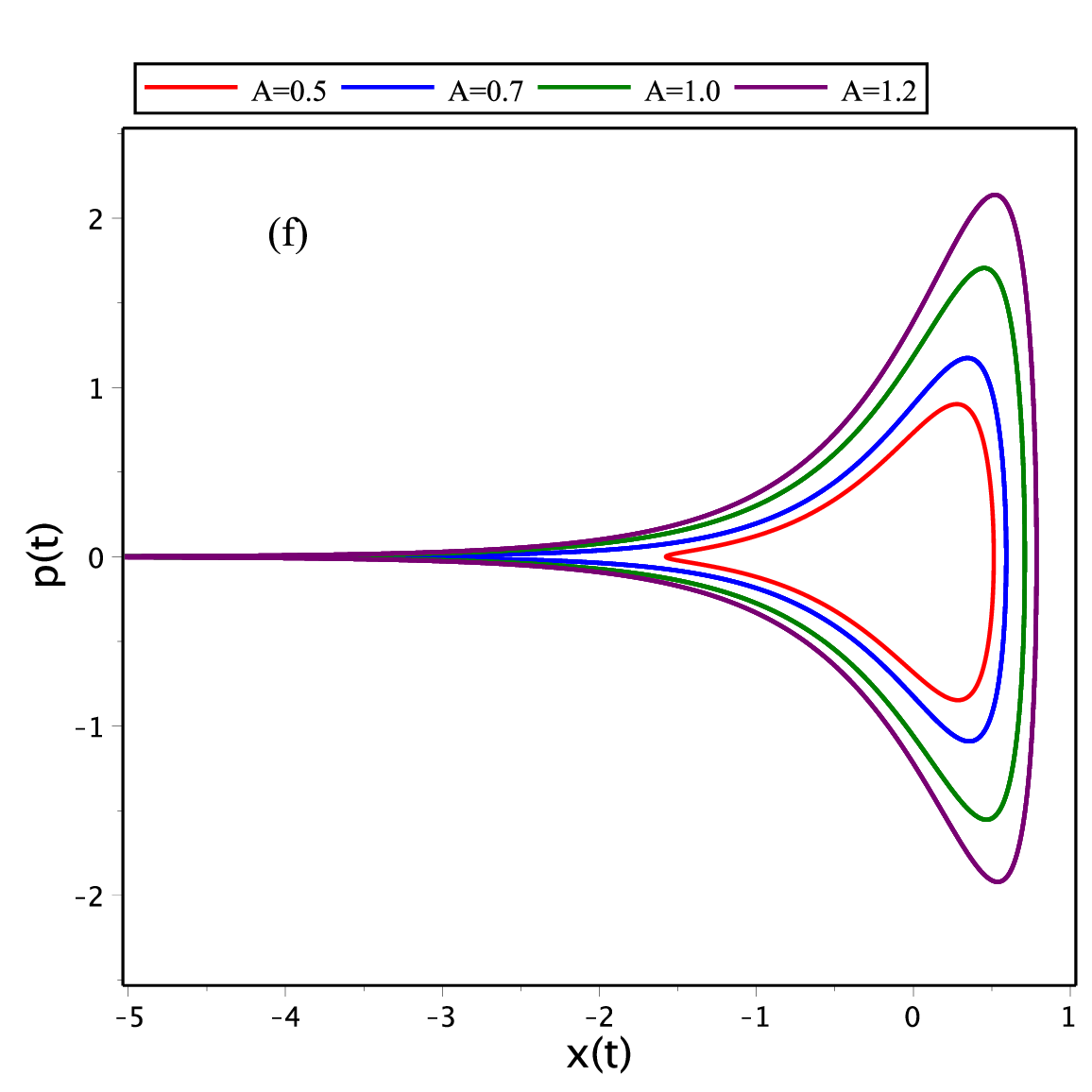}
\caption{\small 
{With $\varphi=0$, we show (a) $x(t)$ of (\ref{x-exponential}) as it evolves in time for different values $\eta$, with $\omega=1$, $A=B=0.25$, and $\lambda=1.5$, (b) $x(t)$ of (\ref{x-exponential}) for different values $\lambda$, with $\omega=1$, $A=B=0.5$, and $\eta=0.05$, (c) $x(t)$ of (\ref{x-exponential}) for different values $A$, with $\omega=1$, $B=0.5$,  $\eta=0.25$, and $\lambda=0.5$. The corresponding phase-space trajectories are shown in (d) for different values of $\eta$ with $A=B=0.25$, $\lambda=1.5$ $\omega=1$, (e) for different values of $\lambda$ with $A=B=0.5$,  $\omega=1$, $\eta=0.05$, and (f) for different values of $A$ with $B=0.5$, $\eta=0.05$, $\lambda=1.2$,$\omega=1$.}}
\label{fig3}
\end{figure}%

\end{document}